\documentclass[mathpazo]{cicp}
\usepackage{amsmath,bm}
\usepackage{subcaption}
\usepackage[pdftex]{graphicx}
\usepackage{hyperref}
\usepackage{xcolor}
\newcommand{\mods}[1]{\textcolor{black}{#1}}

\begin{document}
\title{Entropic multi-relaxation-time lattice Boltzmann model for large density ratio two-phase flows}


\author[Zhang Z R et.~al.]{S.A. Hosseini\affil{1},
      B. Dorschner\affil{1} and I.V. Karlin\affil{1}\comma\corrauth}
\address{\affilnum{1}Department of Mechanical and Process Engineering, ETH Zurich, 8092 Zurich, Switzerland.}
\emails{{\tt shosseini@ethz.ch} (S.A.~Hosseini), {\tt bdorschn@ethz.ch} (B.~Dorschner),
         {\tt ikarlin@ethz.ch} (I.V.~Karlin)}

\begin{abstract}
We propose a multiple relaxation time entropic realization of a two-phase flow lattice Boltzmann model we introduced in earlier works [\href{ 	
https://doi.org/10.48550/arXiv.2112.01975}{S.A. Hosseini, B. Dorschner, and I. V. Karlin, arXiv preprint, arXiv:2112.01975 (2021)}]. 
While the  original model with a single relaxation time allows us to reach large density ratios, it is limited in terms of stability with respect to non-dimensional viscosity and Courant--Friedrichs--Lewy number. 
Here we show that the entropic multiple relaxation time model extends the stability limits of the model significantly, which allows us to reach larger Reynolds numbers for a given grid resolution.
The thermodynamic properties of the solver, using the Peng--Robinson equation of state, are studied first using simple configurations. Co-existence densities and temperature scaling of both the interface thickness and the surface tension are shown to agree well with theory. The model is then used to simulate the impact of a drop onto a thin liquid film with density and viscosity ratios matching those of water and air both in 2-D and 3-D. The results are in very good agreement with theoretically predicted scaling laws and experimental data.
\end{abstract}


\maketitle

\section{Introduction\label{sec:introduction}}
Due to their presence in a wide range of applications, development of models for two-phase flows simulation holds an especially important place in any of the many numerical methods for computational fluid mechanics. The lattice Boltzmann method, developed in the early 90's is no exception to that general observation. Early on after the development of the first lattice Boltzmann models~\cite{mcnamara_use_1988}, extensions to two-phase flow physics were proposed~\cite{gunstensen_lattice_1991,shan_lattice_1993}. Over the past 30 years, a variety of formulations for two-phase flows, from the more classical Allen--Cahn and Cahn--Hilliard based formulation to the very popular pseudo-potential model~\cite{shan_lattice_1993}, have been proposed and widely used. While routinely applied to many different configurations and used to model different physical phenomena, most of these approaches have struggled with large density ratio and high Reynolds number simulations~\cite{chen_critical_2014,li_lattice_2016}. In the class of pseudo-potential and free energy formulations~\cite{swift_lattice_1995,mazloomi_entropic_2015}, these limitations appear in the form of deviations of the coexistence liquid/vapor densities from their analytical counter-parts at lower temperatures and are broadly referred to as thermodynamic inconsistency issues~\cite{chen_critical_2014}.\\
In a recent work we proposed a kinetic scheme and lattice Boltzmann realization exhibiting both thermo- and hydrodynamic consistency even at extremely high density ratios~\cite{hosseini_towards_2021}. It was shown that the scheme was not only suitable to capture thermodynamic properties tied to the liquid-vapor interface and thermodynamically well-posed but also allowed for simulation of dynamic configurations at very high density ratios. However, relying on the simplest collision operator, i.e. single relaxation time, the simulations were limited in terms of the minimum non-dimensional viscosities and the maximum Courant--Friedrichs--Lewy (CFL) numbers that could be achieved, same as for any of the other LB model for two-phase flows available in the literature. As a remedy, more advanced collision operators such as entropic~\cite{ansumali_single_2002,ansumali_stabilization_2000,montessori2017entropic}, multiple relaxation time~\cite{dhumieres_generalized_1992} and regularized~\cite{latt_lattice_2006} have been proposed. The multiple relaxation time collision operator has grown into the most widely used approach, for both single and two-phase flows. While effectively allowing for extended stability domains~\cite{lallemand2000theory,gorban2014enhancement,hosseini2020development,wissocq2020linear,simonis2021linear}, it lacks closures for the individual relaxation rates of the higher order moments.
The entropic multiple relaxation model provides a physically motivated closure for the free parameter and in doing so allows for extended stability domains without tunable parameters~\cite{karlin_gibbs_2014}. A realization for two-phase flows based on a free-energy formulation was devised in \cite{bosch_entropic_2018}. \\
Here, we propose a multiple relaxation time entropic realization of our previously proposed model and thus increase the attainable Reynolds numbers. After a brief introduction of the model, it is first validated on simple configurations which probe thermodynamic properties such as coexistence densities at different temperatures and surface tension. Subsequently, simulations of drop impact on a liquid film are carried out first in two and then in three dimensions, at a density ratio of $10^3$. 
The results show that the entropic multiple relaxation model provides a simple and effective means to overcome the stringent stability limits of the single relaxation time model.
\section{Model description}
The two-phase fluid is modeled using the continuum kinetic framework detailed in~\cite{hosseini_towards_2021} and represented as
\begin{equation}\label{eq:gen_kinetic_model_ext}
	\partial_t f + \bm{v}\cdot\bm{\nabla} f = -\frac{1}{\tau}\left(f - f^{\rm eq}\right) - \frac{1}{\rho}\frac{\partial f^{\rm eq}}{\partial \bm{u}}\cdot\left[\bm{\nabla}\left(P-P_0\right)-\kappa \rho\bm{\nabla}\bm{\nabla}^2\rho\right].
\end{equation}
where $f$ is the one-particle distribution function, $\bm{v}$ the particle velocity, $\kappa$ the capillary coefficient in the second-gradient fluid model, $\rho$ and $\bm{u}$ are the fluid density and velocity, $P$ is the pressure and $P_0$ is a reference pressure used in the equilibrium distribution function $f^{\rm eq}$,
\begin{equation}
	f^{\rm eq}=\frac{\rho}{\left(2\pi P_0/\rho\right)^{D/2}}\exp\left[-\frac{(\bm{v}-\bm{u})^2}{2P_0/\rho}\right].
\label{eq:eq_ref}
\end{equation}
At variance with a classical Boltzmann--Vlasov equation~\cite{vlasov_many-particle_1961}, the Enskog model~\cite{enskog_warmeleitung_1921,chapman_mathematical_1939} or the revised Enskog theory~\cite{van_beijeren_modified_1973},
here the reference pressure $P_0$ is not necessarily the kinetic contribution to the full pressure tensor. Rather, $P_0$ is a parameter that can be adjusted to optimize stability properties of the discrete solver. For the model to be well-posed, $P_0$ must satisfy a sub-isentropic condition,
\begin{equation}\label{eq:subisentropic}
    {P_0}\le C\rho^{5/3},
\end{equation}
for some $C>0$. This guarantees a dissipative evolution equation with respect to normal modes~\cite{hosseini_towards_2021}.\\
Upon discretization in both space and time over $\delta t$, and using the exact difference method~\cite{kupershtokh_new_2004} to evaluate the body force term, the LBGK equations are written,
\begin{equation}\label{eq:LBGK}
	f_i\left(\bm{r}+\bm{c}_i \delta t, t+\delta t\right) = \left(1-\frac{\omega}{2}\right)f_i\left(\bm{r}, t\right) + \frac{\omega}{2} f^{\rm mirr}_i\left(\bm{r}, t\right) + \left(f_i^*-f_i^{\rm eq}\right),
\end{equation}
where $f_i$ and $\bm{c}_i$ are the discrete populations and corresponding discrete particle velocities, $\omega$ \mods{is} the relaxation rate \mods{tied to the fluid kinematic viscosity $\nu$ as:}
\begin{equation}
    \omega = \frac{\delta t}{\rho\nu/P_0 + \delta t/2},
\end{equation}
\mods{and $\delta t$ and $\delta r$ are the time-step and grid spacing.} \mods{Moreover,} $f^{\rm eq}_i$ and $f^*_i$ \mods{are respectively the} discrete equilibrium and extended equilibrium populations to be defined in the next paragraphs. \mods{Finally,} the mirror state \mods{$f^{\rm mirr}_i$} for a single-relaxation time operator \mods{is defined as}~\cite{karlin_gibbs_2014},
\begin{equation}
        f^{\rm mirr}_i\left(\bm{r}, t\right) =  2 f^{\rm eq} - f_i\left(\bm{r}, t\right).
\end{equation}
\subsection{Standard lattice and product-form}
The model is realized on the standard discrete velocity sets $D3Q27$ and $D2Q9$ where $D=2,3$ stands for the spatial dimensions and $Q=9,27$ is the number of discrete velocities. For the sake of clarity the more general case of the $D3Q27$ will be used for the presentation,
\begin{equation}\label{eq:d3q27vel}
	\bm{c}_i=(c_{ix},c_{iy},c_{iz}),\ c_{i\alpha}\in\{-1,0,1\}.
\end{equation}
We first introduce the triplet of functions in the two variables $\xi_{\alpha}$ and $\zeta_{\alpha\alpha}$ that will be defined later, 
\begin{align}
	&	\Psi_{0}(\xi_{\alpha},\zeta_{\alpha\alpha}) = 1 - \zeta_{\alpha\alpha}, 
	\label{eqn:phi0}
	\\
	&	\Psi_{1}(\xi_{\alpha},\zeta_{\alpha\alpha}) = \frac{\xi_{\alpha} + \zeta_{\alpha\alpha}}{2},
	\label{eqn:phiPlus}
	\\
	&	\Psi_{-1}(\xi_{\alpha},\zeta_{\alpha\alpha}) = \frac{-\xi_{\alpha} + \zeta_{\alpha\alpha}}{2},
	\label{eqn:phis}
\end{align}
and a product-form associated with the discrete velocities $\bm{c}_i$ (\ref{eq:d3q27vel}),
\begin{equation}\label{eq:prod}
    	\Psi_i= \Psi_{c_{ix}}(\xi_x,\zeta_{xx}) \Psi_{c_{iy}}(\xi_y,\zeta_{yy}) \Psi_{c_{iz}}(\xi_z,\zeta_{zz}).
\end{equation}
All populations are to be determined by specifying the variables $\xi_\alpha$ and $\zeta_{\alpha\alpha}$ in the product-form (\ref{eq:prod}). 
For the remainder of this work we use $\delta\bm{r}_{i}=\bm{c}_i\delta t$ for the lattice links, and represent the grid spacing in all directions $\alpha=x,y,z$ as $\delta r=\vert c_{i\alpha}\vert\delta t$, $c_{i\alpha}\ne 0$.
\subsection{Discrete equilibrium and extended-equilibrium functions}
Following ~\cite{hosseini_towards_2021} and using the reference pressure $P_0$ the variables appearing in the triplet functions are set to:
\begin{align}
    &\xi_{\alpha}=u_{\alpha},\\
    &\zeta_{\alpha\alpha}=\frac{P_0}{\rho}+u_{\alpha}^2,
\end{align} 
where the local density $\rho$ and  flow velocity $\bm{u}$ are computed as:
\begin{align}
    &	\rho(\bm{r},t)=\sum_{i=0}^{Q-1}f_i(\bm{r},t),\\
    &	\rho\bm{u}(\bm{r},t)=\sum_{i=0}^{Q-1}\bm{c}_if_i(\bm{r},t).
\end{align}
The local equilibrium populations are computed with the product-form \eqref{eq:prod},
\begin{equation}\label{eq:LBMeq}
        f_i^{\rm eq}=
        \rho\prod_{\alpha=x,y,z}\Psi_{c_{i\alpha}}\left(u_\alpha,\frac{P_0}{\rho}+u_{\alpha}^2\right).
\end{equation}
The extended equilibrium populations $f_i^*$ are similarly represented by the product-form (\ref{eq:prod}) and re-defining the variables,
\begin{align}
	&\xi_{\alpha}^{*} = u_{\alpha}+\frac{F_{\alpha}\delta t}{\rho},\label{eq:xistar}	\\
	  &\zeta_{\alpha\alpha}^{*} = \frac{P_0}{\rho} +u_{\alpha}^2 + \frac{\Phi_{\alpha\alpha}}{\rho},\label{eq:zetastar}
\end{align}
where $\Phi_{\alpha\alpha}/\rho$ is the correction term for the diagonals of the non-equilibrium momentum flux tensor~\cite{saadat_extended_2021},
\begin{equation}\label{eq:correction}
	   \Phi_{\alpha\alpha} = \left(1-\frac{\omega}{2}\right) \delta t\partial_{\alpha}\left(\rho u_{\alpha} \left(u_{\alpha}^2 + \frac{3P_{0}}{\rho}-3\varsigma^2\right)\right),
\end{equation}
where we have introduced the so-called lattice speed of sound $\varsigma= \delta r/\sqrt{3}\delta t$. The extended equilibrium is therefor written as,
\begin{equation}\label{eq:LBMstar}
	f_i^*=\rho\prod_{\alpha=x,y,z}\Psi_{c_{i\alpha}}\left(u_\alpha+\frac{F_{\alpha}\delta t}{\rho},\frac{P_0}{\rho}+u_{\alpha}^2+\frac{\Phi_{\alpha\alpha}}{\rho}\right).
\end{equation}
\subsection{Pseudo-potential and capillarity}
In order to recover the correct Korteweg stress tensor~\cite{korteweg_sur_1901} the force is defined as:
\begin{align}
	    \delta t\bm{F} = &\pm 2\psi(\bm{r})\sum_{i=0}^{Q-1} \frac{w_i}{\varsigma^2}  \bm{c}_i \left[\frac{4}{3}\psi(\bm{r}+\bm{c}_i\delta t) - \frac{1}{6} \psi(\bm{r}+2\bm{c}_i\delta t)\right]\nonumber\\
	    &+\tilde{\kappa}\rho(\bm{r})\sum_{i=0}^{Q-1} \frac{w_i}{\varsigma^2} \bm{c}_i \left[2\rho(\bm{r}+\bm{c}_i\delta t) - \rho(\bm{r}+2\bm{c}_i\delta t)\right] + {O}\left([\delta r\bm{\nabla}]^5\right),
	    \label{eq:Kforce_final}
\end{align}
where \mods{$\tilde{\kappa}=\kappa\delta r^2$ and} the pseudo-potential $\psi$ is introduced, 
\begin{equation}
\psi= 
\begin{cases}
    \sqrt{ P-P_0}, & \text{if } P>P_0,\\
    \sqrt{P_0-P}, & \text{if } P \leq P_0,
\end{cases}
\end{equation}
and the weights $w_i$ are defined by the product-form (\ref{eq:prod}) at $\xi_{\alpha}=0$, $\zeta_{\alpha\alpha}=\varsigma^2$, 
\begin{equation}\label{eq:wi}
    w_i=\prod_{\alpha=x,y,z}\Psi_{c_{i\alpha}}\left(0,\varsigma^2\right).
\end{equation}
The square-root form was initially proposed in~\cite{shan_simulation_1994} in an attempt to match the pressure in the pseudo-potential model~\cite{shan_lattice_1993} with the Enskog equation and later reprised in~\cite{yuan_equations_2006} as a way to introduce generic equations of state into the pseudo-potential formulation. As noted previously~\cite{shan_simulation_1994,he_thermodynamic_2002}, while successfully introducing different equations of state into the model, the divergence of the full stress tensor is  different from 
Korteweg's stress tensor as required by the standard thermodynamics~\cite{anderson_diffuse-interface_1998}. While the present model recovers a surface tension term of the form $\kappa\rho\bm{\nabla} {\nabla}^2\rho$ in the continuum limit, 
matching exactly the Korteweg's contribution, the former get\mods{s} a  different term, i.e. $\kappa\psi\bm{\nabla} {\nabla}^2\psi$~\cite{sbragaglia_continuum_2009,sbragaglia_consistent_2011}. Below for optimal performance we set $P_0/\rho=\varsigma^2$.
\subsection{Multi-relaxation time entropic realization}
The two-relaxation time entropic formulation is realized by writing discrete populations as \cite{karlin_gibbs_2014}:
\begin{equation}
        f_i = k_i + s_i + h_i.
\end{equation}
where the kinematic part $k_i$, represents contributions from conserved moments, $s_i$ contributions from the stress and $h_i$ all higher-order moments contributions. Considering invariance of conserved moments and physical constraint on the relaxation rate of second-order moments defining $s_i$, the mirror state can be written:
\begin{equation}
        f_i^{\rm mirr} = k_i + \left(2s_i^{\rm eq} - s_i\right) + (1-\gamma)h_i + \gamma h^{\rm eq},
\end{equation}
where a free parameter $\gamma$ has been introduced, which allows independent control over the relaxation rate of higher-order moments. This free parameter is found by minimizing the discrete entropy in the post-collision state, $f'_i$:
\begin{equation}
    \frac{dH(f')}{d\gamma} = 0,
\end{equation}
which upon expansion around equilibrium up to the first non-vanishing order results in~\cite{bosch_entropic_2018}:
\begin{equation}
    \frac{\gamma}{2} = \frac{1}{\omega}-\left(1-\frac{1}{\omega}\right)\frac{\langle\Delta s\lvert \Delta h\rangle}{\langle\Delta h\lvert \Delta h\rangle},
\end{equation}
where $\Delta s_i= s_i^{\rm eq}-s_i$, $\Delta h_i= h_i^{\rm eq}-h_i$, and the entropic scalar product $\langle\lvert\rangle$ is defined as:
\begin{equation}
    \langle X\lvert Y\rangle=\sum_{i=1}^{Q}\frac{X_i Y_i}{f^*_i}.
\end{equation}
Here we use the central Hermite moments as basis for the projection. Details of the moments space and corresponding contributions are given in Appendix \ref{app:moments_space}. \mods{Furthermore, a set of Matlab codes and functions to derive analytical expressions for the proposed collision operator are available in supplementary materials.}
\section{Numerical validation}
The model as introduced in the previous section will be used here to conduct both thermodynamic checks and dynamic simulations.
\subsection{Consistency tests}
We first validate the thermodynamic consistency of the solver for the density ratios of interest, here $\approx10^3$. Simulation are performed using the Peng-Robinson equation of state \cite{peng_new_1976},
\begin{equation}\label{eq:PREoS}
    P = \frac{\rho R T}{1-b\rho} - \frac{a \alpha(T) \rho^2}{1+2\rho b - b^2 \rho^2},
\end{equation}
with
\begin{equation}
    \alpha(T) = \left[1 + (0.37464 + 1.54226\omega' - 0.26992 \omega'^2) \left(1-\sqrt{T/T_c}\right)\right]^2,
\end{equation}
where  $\omega'$ is the acentric factor ($\omega'=0.344$ for water), and
\begin{equation}
    a=0.45724\frac{R^2T_c^2}{P_c},\ b=0.0778\frac{R T_c}{P_c}.
\end{equation}
\mods{Here $T_c$ and $P_c$ are the critical state temperature and pressure, $a$ and $b$ are constants accounting for the strength of the attractive inter-molecular force and the volume occupied by molecules, and $R$ the universal gas constant.}
\subsubsection{Co-existence densities}
\begin{figure}[!h]
\centering
	 \includegraphics[width=0.9\columnwidth]{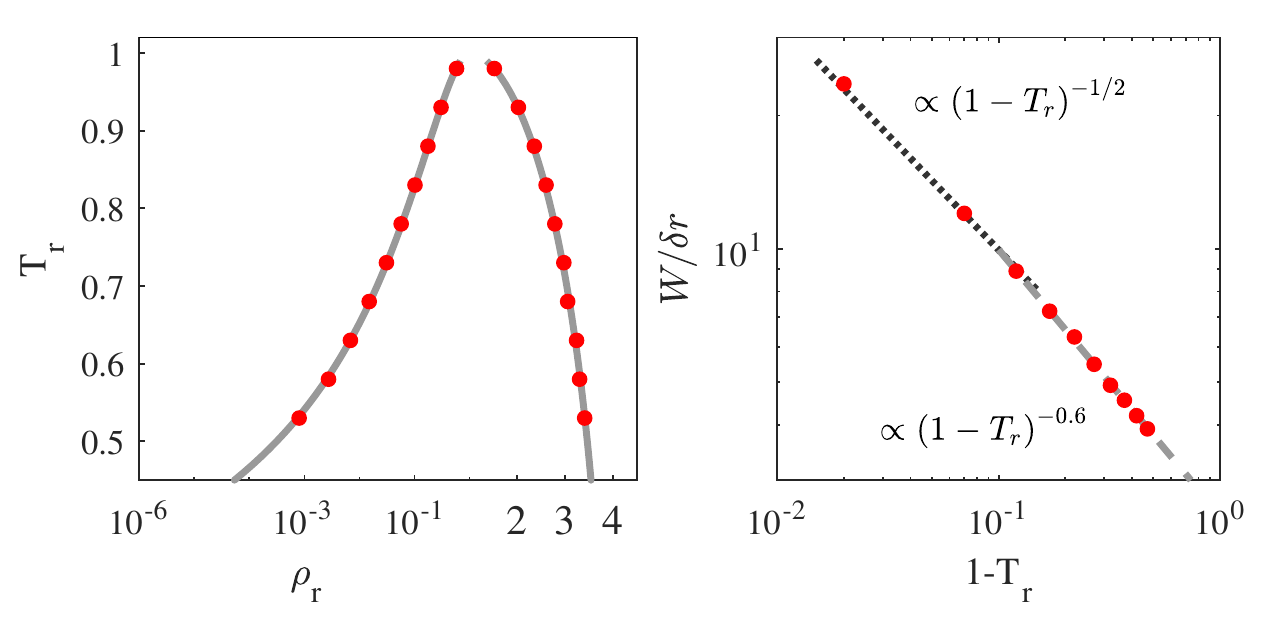}
    \caption{Left: Coexistence densities for the Peng-Robinson equation of state as obtained from (grey line) the Maxwell construction and (red markers) numerical simulations with $a=0.003$ and $b=0.095$. Right: Interface thickness for different temperatures (red markers) as obtained from simulations. 
    }
\label{Fig:Maxwell_Thickness}
\end{figure}

We begin with the validation of liquid-vapour coexistence. Two-dimensional flat interface simulations are conducted on $800\times10$ grid-points. The domain is filled with the vapour phase of the fluid and periodic boundary conditions are applied all around. A 400 points wide column of the liquid phase is placed at the center. Simulations are ran until steady-state, characterized with a $L_0$ norm convergence criterion based on the density field, is reached. The theoretical prediction for the coexistence density ratio $\rho_l/\rho_v$ can be obtained via the equilibrium condition leading to the Maxwell equal-area construction,
\begin{equation}\label{eq:Maxwell_equal_area}
    \int_{\rho_v}^{\rho_l} \frac{P_{\rm sat}-P}{\rho^2}d\rho=0,
\end{equation}
where $P_{\rm sat}(T)$ is the saturation pressure at which the liquid and vapor phases coexist at a given temperature $T$ below the critical point. The results as obtained from both theory and simulation are shown in Fig.~\ref{Fig:Maxwell_Thickness} and point to an excellent agreement. This is not surprising as the interface thickness, defined here as $W=(\rho_l-\rho_v)/{\rm max}\lvert\bm{\nabla}\rho\lvert$, points to well-resolved interfaces down to the temperature of interest as shown in Fig.~\ref{Fig:Maxwell_Thickness}.
\subsubsection{Surface tension}
\begin{figure}[!h]
	\centering
		 \includegraphics[width=0.9\columnwidth]{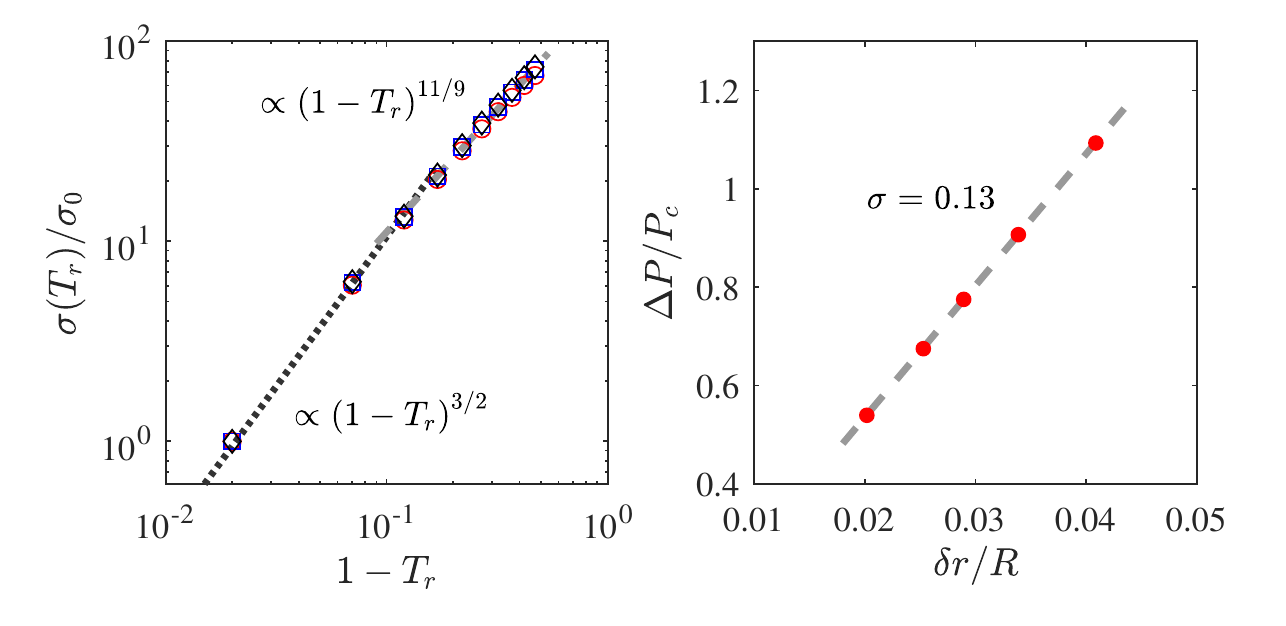}
\caption{Left: 
Surface tension for different temperatures and values of $a$, (red circles) $a=0.003$, (blue squares) $a=0.0017$ and (black diamonds) $a=0.0011$ with $b=0.095$. Right: Illustration of the Laplace law for $T_r$=0.59 and $a=0.003$.}
\label{Fig:Surf_Tension}
\end{figure}

\begin{figure}[!h]
	\centering
		 \includegraphics[width=0.9\columnwidth]{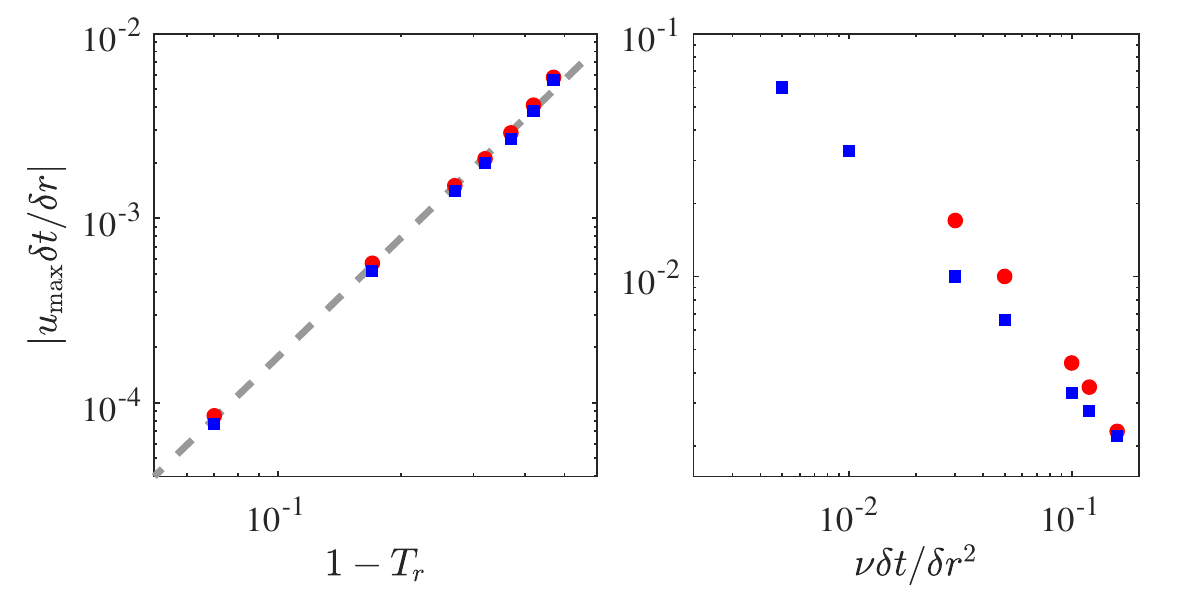}
    \caption{Maximum spurious currents for (left) different temperatures and $\nu\delta t/\delta r^2=0.1$ and (right) viscosities at $T_r=0.59$ as obtained from 2-D drop simulations with (red circular markers) single relaxation time and (square blue markers) entropic collision operators.}
\label{Fig:spurious_currents}
\end{figure}

\begin{figure}[!h]
	\centering
		 \includegraphics[width=0.4\columnwidth]{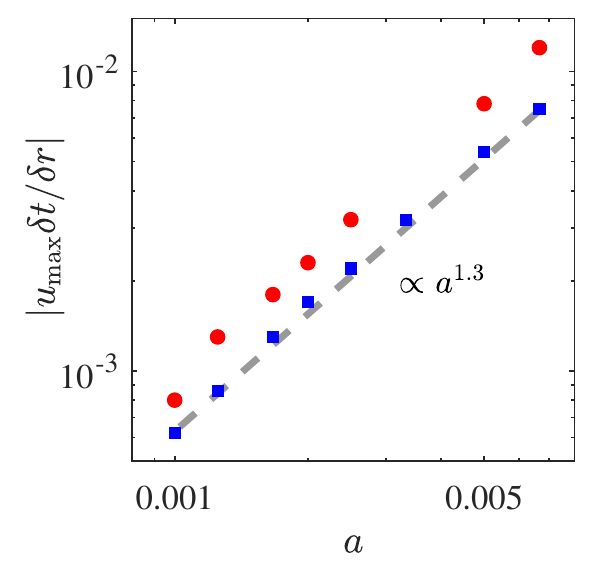}
    \caption{Maximum spurious currents for different values of $a$ at $T_r=0.59$ \mods{with (red circular markers) single relaxation time and (square blue markers) entropic collision operators}.}
\label{Fig:spurious_currents_a}
\end{figure}

The meanfield behavior of the van der Waals second gradient fluid is known to lead to a scaling with temperature of the form
\begin{equation}\label{eq:vdw_surface_tension}
    \sigma \propto {\left(1-T_r\right)}^{3/2}.
\end{equation}
The scaling has been theoretically demonstrated by van der Waals close to the critical temperature $T_r\rightarrow 1$~\cite{van_der_waals_thermodynamische_1894,blokhuis_thermodynamic_2006}. To further validate the thermodynamic properties of the solver 2-D simulations of drops with initial radii $R_0/\delta r\in\left[25, 200\right]$ with $R_0/W > 10$ placed at the center of a fully periodic square domain are considered. The surface tension coefficient is evaluated using the Laplace law ($D=2$) as, 
\begin{equation}\label{eq:Laplace_equimolar}
    \Delta P = \frac{(D-1)\sigma}{R_e},
\end{equation} 
where the radius $R_e$ is the \emph{equimolar} dividing surface \cite{gibbs_equilibrium_1874}. Simulations were conducted for different temperatures with $T_r\in[0.5,\, 0.98]$ and three different values of $a$. The obtained results are illustrated in Fig.~\ref{Fig:Surf_Tension}.
In agreement with the van der Waals model the surface tension scales as $3/2$ near the critical point. Below $T_r=0.8$ the scaling changes slightly to $11/9$ which is in agreement with the fit proposed by Guggenheim~\cite{guggenheim_principle_1945}. Furthermore, in agreement with the theory of corresponding state proposed by Guggenheim, these scalings hold regardless of the choice of value for the interaction parameters~\cite{guggenheim_principle_1945}. For simulations however, as demonstrated in our previous publication the principle of corresponding states is only correctly recovered in the limit of a resolved interface, i.e. $\delta r/W\rightarrow 0$, that we characterized as \emph{thermodynamically converged} simulations~\cite{hosseini_towards_2021}.\\
Additionally the maximum spurious currents for each one of the considered temperatures and different non-dimensional viscosities were monitored for the same choice of coefficients in the equation of state using both the single relaxation time and entropic collision operators. The results are shown in Fig.~\ref{Fig:spurious_currents}.
While the use of the entropic collision operator has very limited effects on the spurious currents for $\nu\delta t/\delta r^2=0.1$, it is observed to reduced them by up to a factor of two at lower viscosities. Furthermore as observed in the plots it extends the domain of stability reaching non-dimensional viscosities as low as $10^{-3}$ \mods{and considerably accelerates the convergence of spurious velocities to their equilibrium values.}
\begin{figure}[!h]
	\centering
		 \includegraphics[width=0.55\columnwidth]{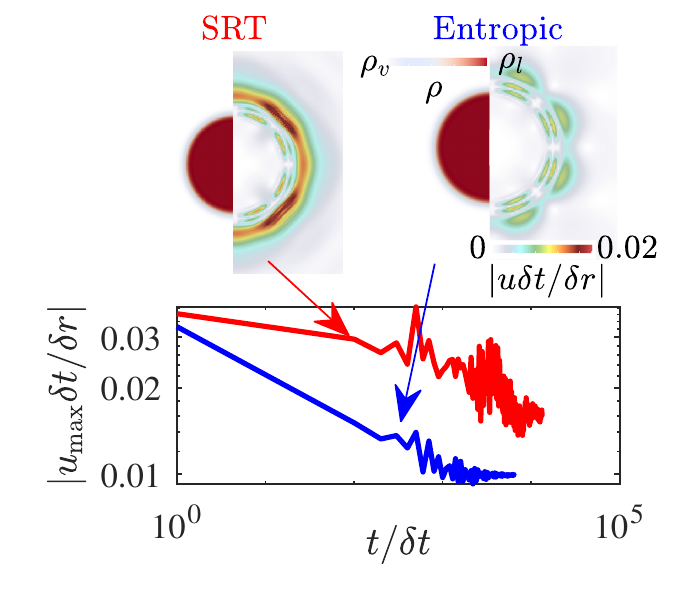}
    \caption{ \mods{(right) Time-evolution of maximum spurious currents along with the density and velocity fields at the converged state as obtained from simulations with SRT and entropic models at $\nu=0.03$ and $T_r=0.59$}.}
\label{Fig:spurious_currents_field}
\end{figure}
\mods{The latter, illustrated in Fig.~\ref{Fig:spurious_currents_field}, can be explained by the fact that putting bulk viscosity under entropy control enhances dissipation of normal modes.} It is also interesting to note that the spurious currents can be fitted with a function of the form ${(1-T_r)}^m$ leading to a scaling of $m\approx2.2$. Spurious currents can also be mitigated by increasing the thickness of the interface. Rescaling of the force contribution via the equation of state coefficient $a$ has been shown to allow for control over the interface thickness. To that end the effect of the choice of that parameter on spurious currents at the temperature of interest has also been studied. The results are shown in Fig.~\ref{Fig:spurious_currents_a} and show that interface thickness can effectively be used to control maximum spurious currents.
\subsection{Crown radius evolution scaling upon impact: 2-D simulation}
Next we consider the impact of a drop on a thin liquid layer in 2-D. It is an interesting configuration as it involves complex dynamics. In many instances, upon impact and at the contact point (line) a thin liquid jet also referred to as \emph{ejecta} is formed, which continues to grow and propagate as a \emph{corolla}
~\cite{marcotte_ejecta_2019}. Detailed studies of the initial stages of the spreading of the crown have shown that the spreading radius scales as the square root of time regardless of the Weber and Reynolds numbers~\cite{josserand_droplet_2003}. To validate the solver, we consider four different sets of Weber and Reynolds numbers, 
\begin{equation}
    ({\rm We}, {\rm Re})\in\{(57,4000),(93,4000),(238,4000),(93,100)\},
\end{equation}
where the non-dimensional parameters are defined as
\begin{equation}
    {\rm Re}=\frac{2\rho_l U_0 R_0}{\mu_l},
\end{equation}
and
\begin{equation}
    {\rm We} = \frac{2\rho_l R_0 {U_0}^2}{\sigma},
\end{equation}
with $R_0$ and $U_0$ the initial drop radius and velocity. The domain is rectangular of size $24R_0\times 12R_0$ with a liquid film of thickness $h=R_0$ at the bottom. In all simulations presented in this section $T_r=0.59$ leading to $\rho_l/\rho_v=10^3$ and $\nu_v/\nu_l=15$ corresponding to the air/water system. The drop radius is set to $R_0=100\delta r$.

\begin{figure}[!h]
	\centering
		 \includegraphics[width=\columnwidth]{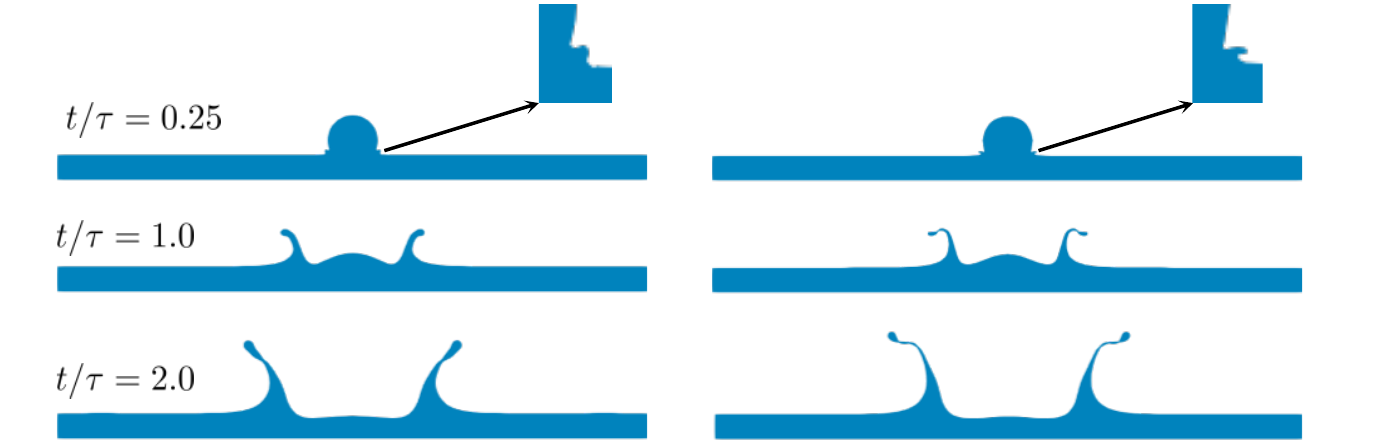}
    \caption{Snapshots of the 2-D drop impacting thin liquid film for (left) We=57, Re=4000 and (right) We=238, Re=4000.}
    \label{Fig:We_effect_evolution}
    \end{figure}

\begin{figure}[!h]
	\centering
		 \includegraphics[width=\columnwidth]{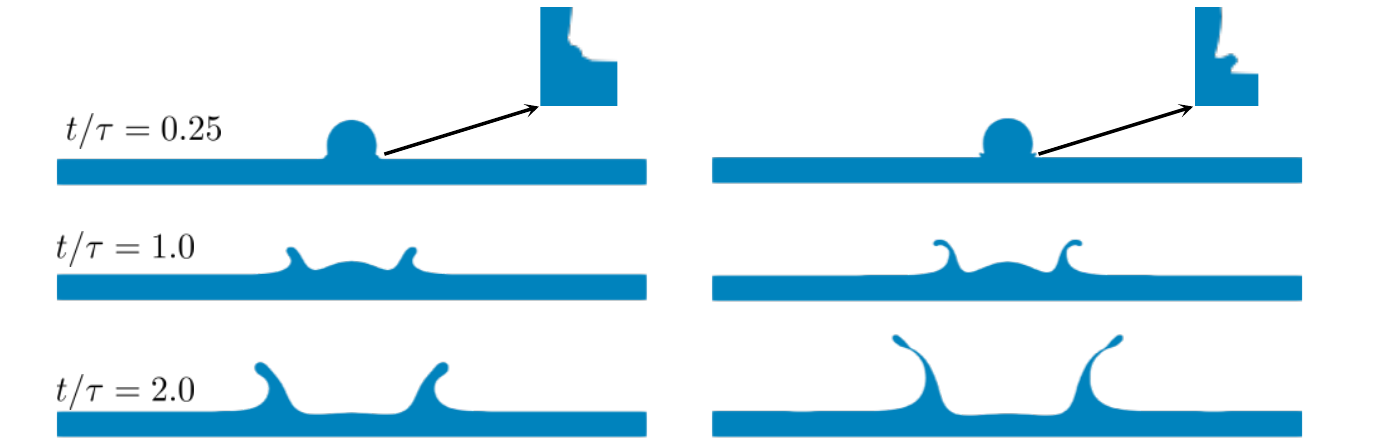}
    \caption{Snapshots of the 2-D drop impacting thin liquid film for (left) We=93, Re=100 and (right) We=93, Re=4000.}
    \label{Fig:Re_effect_evolution}
    \end{figure}

\begin{figure}[!h]
	\centering
		 \includegraphics[width=0.45\columnwidth]{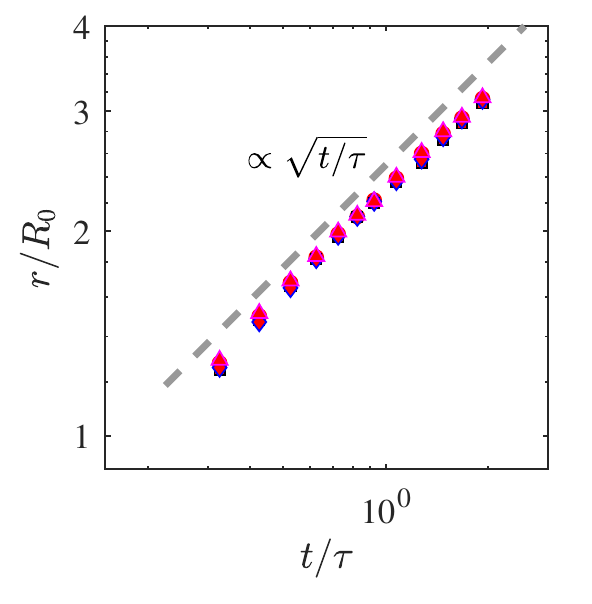}
    \caption{Evolution of the crown radius over time as obtained from 2-D simulation. The dashed line represents the $\propto \sqrt{t/\tau}$ scaling.}
    \label{Fig:2D_crown_radii}
    \end{figure}

\begin{figure}[!h]
	\centering
		 \includegraphics[width=0.45\columnwidth]{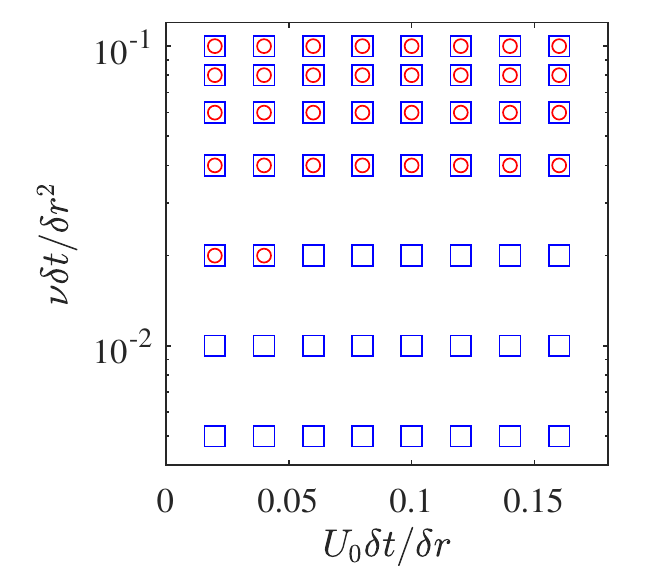}
\caption{\mods{Stability domain as obtained for the 2-D drop impact simulation as a function of non-dimensional impact velocity and viscosity. The drop radius is set to $R_0=50\delta r$. Single relaxation time collision operator stability domain are shown with red circular symbols while those of the entropic model are shown with blue square markers. 
}}
\label{Fig:Stability}
\end{figure}

The evolution of the liquid surface as obtained from the simulations is shown in Figures~\ref{Fig:We_effect_evolution} and \ref{Fig:Re_effect_evolution}. The times are reported in non-dimensional form, normalized by the convective characteristic time $\tau=2R_0/U_0$.
It is interesting to note that at the lowest Reynolds number, Fig.~\ref{Fig:Re_effect_evolution}, the ejecta does not form. Furthermore at this Reynolds number the crown is thicker and smaller in height. All these observations are expected as the larger liquid viscosity leads to faster dissipation of the initial kinetic energy of the drop, hence lower heights reached by the crown, and thicker boundary layers explaining both the thickness of the ejecta and crown.
The evolution of the spreading crown radii $r$ over time for different cases are shown in Figure~\ref{Fig:2D_crown_radii}. As shown there the radii scale as the square root of time at the initial stages of the impact, in agreement with results reported in~\cite{josserand_droplet_2003}.
It should be noted that using the same resolution, i.e. $R_0=100\delta r$, the single relaxation time collision operator was also stable for simulations up to Re=2000. \mods{To better illustrate the effect of the entropic collision operator on stability, simulations were run using different impact speeds and viscosities for a fixed interface thickness $W=5\delta r$. Simulations were carried out with $\nu\delta t/\delta r^2 \in [0.0025\text{ }0.1]$ and $U_0\delta t/\delta r\in[0\text{ }0.16]$.  The resulting stability domains are plotted in Fig.~\ref{Fig:Stability}.As shown there, while the single relaxation solver becomes unstable below $\nu\delta t/\delta r^2=0.04$ the entropic solver remains stable down to $\nu\delta t/\delta r^2=0.005$. Below that value simulations become unstable because spurious currents get too large. This limit can in practice be pushed further down by reducing spurious currents via thicker interfaces.}
\subsection{3-D configuration: comparison with experiments}
\begin{figure}[!h]
	\centering
		 \includegraphics[width=1\columnwidth]{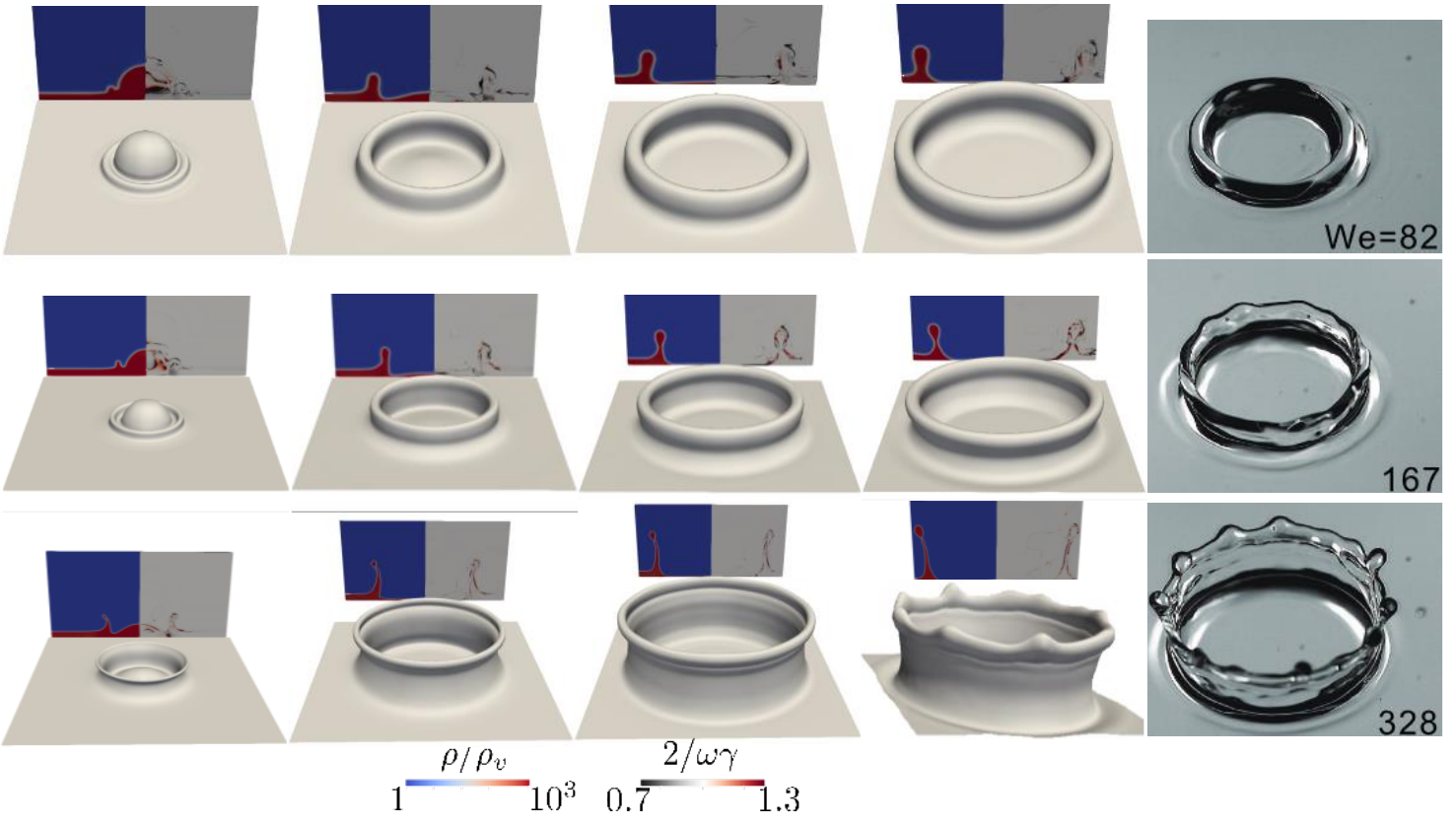}
\caption{Snapshots at (from left to right) t=0.1, 0.3, 0.5 and 0.7~ms of simulations of 3-D drop impact on liquid film. The right-most column shows snapshots at t=0.7~ms from experiments reported in \cite{che_impact_2017}. The rows (from top to bottom) correspond to We=82, 167 and 328. The planes behind iso-surfaces representing the liquid show the distribution of density and $\frac{2}{\omega\gamma}$ on the central plane.
}
\label{Fig:splash_3D_images}
\end{figure}

As a final configuration we consider the impact of a drop on a thin liquid film in 3-D. Configurations follow the experimental study presented in \cite{che_impact_2017}. Three different cases are studied in this section, We=82, 167 and 328 and they cover a Range of Reynolds numbers between 3500 to 6900. All simulations parameters are chosen so as to match a water/air system. Furthermore the thickness of the liquid film $h$ is set to $h/2R_0=0.22$ following \cite{che_impact_2017}. In all simulations the drop initial radius is set to $R_0=80$ and the domain size to $14 R_0\times14 R_0\times6R_0$. \mods{For the largest We number configuration, in order to initiate instability without enforcing any specific wavelength, the velocity field in the liquid phases is initially supplemented with white noise with a normal distribution centered around the initial uniform velocity and a standard deviation of the order of 5 percent of the impact velocity.} The evolution of the crown radius over time is then extracted from simulations and compared to experimental data reported in \cite{che_impact_2017}.\\
The evolution of the liquid surface over time for all three Weber numbers along with the relaxation rate of higher order moments are shown in Fig.~\ref{Fig:splash_3D_images} along with corresponding snapshots from experiments.
It is interesting to note that, as expected, the entropic stabilizer is highly active near liquid/vapor interfaces. In parts of the domain away from interfaces $2/\omega\gamma\approx 1$. At the largest Weber number the snapshot from the experiment shows instabilities on the crown. In the corresponding simulation, while being less pronounced an onset of instability on the crown can clearly be observed. The differences in the amplitude of the instability at a given time between experiments and simulations can be explained by higher levels of perturbation and noise present in typical experimental condition speeding up the growth of unstable modes~\cite{rieber_numerical_1999,cheng_numerical_2015}.
\begin{figure}[!h]
	\centering
		 \includegraphics[width=0.6\columnwidth]{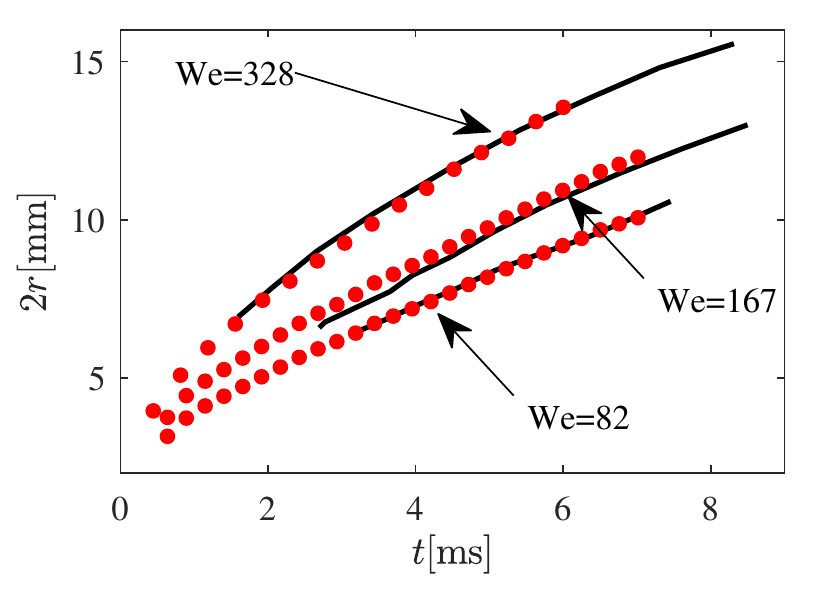}
\caption{Evolution of the crown diameters over time (red circular markers) as obtained from simulation and (black lines) reported in \cite{che_impact_2017} from experiments.}
\label{Fig:impact_3D_radius}
\end{figure}
The changes in the crown diameter as obtained from both simulations and experiments reported in \cite{che_impact_2017} are shown in Fig.~\ref{Fig:impact_3D_radius}. The results point to excellent agreement between experimental observations and numerical simulation conducted with the entropic model. \mods{A video of the drop evolution over time for We=328 can be found in supplementary materials.}
\section{Conclusion}
We proposed a multiple relaxation time entropic realization of a previously introduced model for two-phase flows. The use of this collision operator was shown to drastically impact the stability domain and to a certain extent the maximum spurious currents allowing simulations to reach larger Re numbers, reaching values up to $\approx7000$ in the case of 3-D drop impact simulations. All of this, as evidenced by co-existence densities and surface tension scaling, without sacrificing the thermodynamic consistency of the solver.\\

\section*{Acknowledgments}
This work was supported by European Research Council (ERC) Advanced Grant no. 834763-PonD (S.A.H, B.D. and I.K.) and the Swiss National Science Foundation (SNSF) grant No. 200021-172640 (S.A.H.). Computational resources at the Swiss National Super Computing Center CSCS were provided under grant no. s1066.
\bibliographystyle{plain}
\bibliography{reference.bib}
\appendix
\section{Moment projection space for the relaxation process\label{app:moments_space}}
A modified set of central Hermite moments are used here for the collision operator as projection space for the relaxation process. The modifications concern second-order moments to allow for independent control over the bulk viscosity as standard Hermite polynomials do not allow it. The Hermit coefficients $a_{x^p y^q}$ where for the $D2Q9$ stencil $a_{x^p y^q}\in\{a_0,a_x,a_y,a_{xy},a_{x^2}-a_{y^2},a_{x^2}+a_{y^2},a_{x^2y}, a_{xy^2}, a_{x^2y^2}\}$ are defined as:
\begin{equation}
    a_{x^p y^q} = \sum_{i=0}^{Q-1} \mathcal{H}_{x^p y^q}(\bm{c}_i-\bm{u}) f_i,
\end{equation}
where $\mathcal{H}_{x^p y^q}$ are the corresponding central Hermit polynomials. For the D3Q27 stencil we define Hermite coefficients as:
\begin{equation}
    a_{x^p y^q z^r} = \sum_{i=0}^{Q-1} \mathcal{H}_{x^p y^q z^r}(\bm{c}_i-\bm{u}) f_i,
\end{equation}
and use the following set of moments:
\begin{multline}
    a_{x^p y^q z^r}\in\{a_0,a_x,a_y,a_z,a_{xy},a_{xz},a_{yz}, a_{x^2}-a_{y^2},a_{x^2}-a_{z^2},a_{x^2}+a_{y^2}+a_{z^2},a_{x^2y}, a_{x^2z}, a_{xy^2}, a_{y^2z}, a_{xz^2},\\ a_{yz^2}, a_{x^2y^2},
    a_{x^2y^2}, a_{x^2z^2}, a_{y^2z^2}, a_{x^2yz}, a_{xy^2z}, a_{xyz^2}, a_{xy^2z^2}, a_{x^2yz^2}, a_{x^2y^2z}, a_{x^2y^2z^2} \}.
\end{multline}
Applying the transform to the equilibrium populations, $a^{\rm eq}_0=\rho$ and $a^{\rm eq}_{x^p y^q z^r}=0$ $\forall (x,p,r)\neq(0,0,0)$ which in turn leads to $k_i^{\rm eq} = f_i^{\rm eq}$ and $s_i^{\rm eq} = h_i^{\rm eq} = 0$. For the extended equilibrium $f^*_i$ the central Hermit moments in the case of the D2Q9 stencil are:
    \begin{align}
    \bm{a}^*=&\nonumber\\
	    & \left[\rho, F_x \delta t, F_y\delta t, \frac{F_x F_y\delta t^2}{\rho}, \frac{\delta t^2(F_x^2-F_y^2)+\rho(\Phi_{xx}-\Phi_{yy})}{\rho}, \frac{\delta t^2(F_x^2+F_y^2)+\rho(\Phi_{xx}+\Phi_{yy})}{\rho}, \right. \nonumber\\
	    & \left. \frac{\delta t F_y(\delta t^2F_x^2+\rho\Phi_{xx})}{\rho^2}, \frac{\delta t F_x(\delta t^2F_y^2+\rho\Phi_{yy})}{\rho^2},  \frac{(\delta t^2F_x^2 +\rho\Phi_{xx})\,(\delta t^2 F_y^2 +\rho\Phi_{yy})}{\rho^3} \right].
	\end{align}
while in 3-D for the D3Q27 stencil:
    \begin{align}
    \bm{a}^*=&\nonumber\\
	    & \left[\rho, F_x \delta t, F_y \delta t, F_z \delta t, \frac{F_x F_y\delta t^2}{\rho}, \frac{F_x F_z\delta t^2}{\rho}, \frac{F_y F_z\delta t^2}{\rho}, \frac{\delta t^2(F_x^2 - F_y^2) + \rho(\Phi_{xx} - \Phi_{yy})}{\rho}, \right. \nonumber\\ &\left. \frac{\delta t^2(F_x^2 - F_z^2) + \rho(\Phi_{xx} - \Phi_{zz})}{\rho}, \frac{\delta t^2(F_x^2 + F_y^2 + F_z^2) + \rho(\Phi_{xx} + \Phi_{yy} + \Phi_{zz})}{\rho}, \frac{\delta t F_y(\delta t^2F_x^2 + \rho\Phi_{xx})}{\rho^2}, \right. \nonumber\\ &\left. \frac{\delta t F_z(\delta t^2F_x^2 + \rho\Phi_{xx})}{\rho^2}, \frac{\delta t F_x(\delta t^2F_y^2 + \rho\Phi_{yy})}{\rho^2}, \frac{\delta t F_z(\delta t^2F_y^2 + \rho\Phi_{yy})}{\rho^2}, \frac{\delta t F_x(\delta t^2F_z^2 + \rho\Phi_{zz})}{\rho^2}, \right. \nonumber\\ &\left.
	    \frac{\delta t F_y(\delta t^2F_z^2 + \rho\Phi_{zz})}{\rho^2}, \frac{\delta t^3 F_x F_y F_z}{\rho^2}, \frac{(\delta t^2F_x^2 + \rho\Phi_{xx}) (\delta t^2F_y^2 + \rho\Phi_{yy})}{\rho^3}, \frac{(\delta t^2F_x^2 + \rho\Phi_{xx}) (\delta t^2F_z^2 + \rho\Phi_{zz})}{\rho^3}, \right. \nonumber\\ &\left.
	    \frac{(\delta t^2F_y^2 + \rho\Phi_{yy}) (\delta t^2F_z^2 + \rho\Phi_{zz})}{\rho^3}, \frac{\delta t^2 F_y F_z (\delta t^2 F_x^2 + \rho \Phi_{xx})}{\rho^3}, \frac{\delta t^2 F_x F_z (\delta t^2 F_y^2 + \rho \Phi_{yy})}{\rho^3}, \right. \nonumber\\ &\left.
	    \frac{\delta t^2 F_x F_y (\delta t^2 F_z^2 + \rho \Phi_{zz})}{\rho^3}, \frac{\delta tF_z (\delta t^2 F_x^2 + \rho\Phi_{xx})(\delta t^2 F_y^2 + \rho\Phi_{yy})}{\rho^4}, \frac{\delta tF_y (\delta t^2 F_x^2 + \rho\Phi_{xx})(\delta t^2 F_z^2 + \rho\Phi_{zz})}{\rho^4}, \right. \nonumber\\ &\left. \frac{\delta tF_x (\delta t^2 F_y^2 + \rho\Phi_{yy})(\delta t^2 F_z^2 + \rho\Phi_{zz})}{\rho^4}, \frac{(\delta t^2 F_x^2 + \rho\Phi_{xx})(\delta t^2 F_y^2 + \rho\Phi_{yy})(\delta t^2 F_z^2 + \rho\Phi_{zz})}{\rho^5} \right].
	\end{align}
\end{document}